\documentclass[twocolumn,aps,pre,preprintnumbers,amsmath,amssymb,nofootinbib,floatfix]{revtex4-1}

\usepackage{graphicx,color}
\usepackage{amsmath}
\usepackage{epstopdf}
\usepackage{soul}
\usepackage{tabularx}

\newcommand{\D}{\mathrm{d}}
\newcommand{\e}{\mathrm{e}}

\newcommand{\eps}{\varepsilon}

\newcommand{\kbt}{k_{\mathrm{B}}T}

\newcommand{\lb}{l_{{\rm B}}}
\newcommand{\ld}{\lambda_{{\rm D}}}

\newcommand{\vecr}{\boldsymbol{r}}
\newcommand{\veck}{\boldsymbol{k}}

\newcommand{\ra}[1]{\textcolor{black}{#1} } % comments in MAGENTA
\newcommand{\raa}[1]{\textcolor{black}{#1} } % comments in blue

\newcommand{\bu} {\boldsymbol{u}}
\newcommand{\bv} {\boldsymbol{v}}
\newcommand{\buo} {\boldsymbol{u}_0}
\newcommand{\bjn} {\boldsymbol{J}}
\newcommand{\bjc} {\boldsymbol{j}}

\newcommand{\bF} {\boldsymbol{F}}
\newcommand{\bz} {\boldsymbol{z}}

\newcommand {\KC} {\mathcal{K}}

\begin{document}
%%%%%%%%%%%%%%%%

%%%%%%%%%%%%%%%%%%%%%%%%%%%%%%%%%%%%%%%%%%%%%%%
\title{Linear response functions of an electrolyte solution in a uniform flow}

\author{Ram M. Adar}
\affiliation{Raymond and Beverly Sackler School of Physics and Astronomy, Tel Aviv
University, Ramat Aviv, Tel Aviv 69978, Israel}
\affiliation{
Department of Chemistry, Graduate School of Science,
Tokyo Metropolitan University, Tokyo 192-0397, Japan}

\author{Yuki Uematsu}
\affiliation{Department of Chemistry, Kyushu University, Fukuoka 819-0395, Japan}

\author{Shigeyuki Komura}\email{komura@tmu.ac.jp}
\affiliation{
Department of Chemistry, Graduate School of Science,
Tokyo Metropolitan University, Tokyo 192-0397, Japan}

\author{David Andelman}\email{andelman@post.tau.ac.il}
\affiliation{Raymond and Beverly Sackler School of Physics and Astronomy, Tel Aviv
University, Ramat Aviv, Tel Aviv 69978, Israel}

%\date{\today}
%\date{Ver 15 -- June 17, 2018 -- Shige}

%\baselineskip=22pt

%%%%%%%%%%%%%%%%%%%%%%%
\begin{abstract}
We study the steady state response of a dilute monovalent electrolyte solution to an external source with a constant relative velocity with respect to the fluid. The source is taken as a combination of three perturbations: an external force acting on the fluid, an externally imposed ionic chemical potential, and an external charge density. The linear response functions are obtained analytically and can be decoupled into three independent terms, corresponding to (i) fluid flow and pressure, (ii) total ionic number density and current, and (iii) charge density, electrostatic potential and electric current. It is shown how the uniform flow breaks the equilibrium radial symmetry of the response functions, leading to a distortion of the ionic cloud and electrostatic potential, which deviate from the standard Debye-H\"uckel result. The potential of a moving charge is under-screened in its direction of motion and over-screened in the opposite direction and normal plane. As a result, an unscreened dipolar electric field and electric currents are induced far from the charged source. We relate our general formalism to several experimental setups, \ra {such as colloidal sedimentation}.
\end{abstract}
%%%%%%%%%%%%%%%%%%%%%%

%%%%%%
\maketitle
%%%%%%

%%%%%%%%%%%%
\section{Introduction}
\label{sec1}

Ionic solutions are found in a wide range of biological systems and are used for a plethora of industrial applications and processes. For ionic solutions out of thermodynamic equilibrium, many well-known physical phenomena rely on the interplay between Coulombic interactions, hydrodynamics, and thermal diffusion~\cite{DoiBook,Dukhin,Masliyah,Shaw}. For example, an applied electric field can induce an electrolyte flow in a capillary
(electrosmosis)~\cite{Bazant04,Squires04,BaznatSquiers04,Loly04,Loly06,Ajdari06,Yuki13}.
Similarly, a salt concentration gradient in a capillary causes a water flow (diffusio-osmosis)~\cite{Anderson1989,Keh2007} and electric currents (osmotic current)~\cite{Anderson1989,Siria2013}.

For colloidal suspensions, transport of charged colloids can be achieved by applying an external electric field (electrophoresis)~\cite{Obrien,Maduar,Kim06,Lobaskin07,Giupponi11}. The distortion of the electric double layer near the colloidal surface\ra{, in the presence of the applied field,} significantly decreases the electrophoretic mobility for high surface potentials~\cite{Obrien}. Moreover, when charged colloids sediment under gravity in an electrolyte solution, a potential difference builds along the direction of gravity, called sedimentation potential~\cite{Ohshima,Booth}.

Electrokinetic processes can be described by the linear response theory of the system to external sources\ra{, when the sources are sufficiently weak}. The response in the quiescent state is well-known when the hydrodynamics and electrostatics are completely decoupled. In such cases, the linear response of the velocity field with respect to a point force is described by the Oseen tensor~\cite{Oseen,Happel}, and that of the electrostatic potential to a point charge is described by the Debye-H\"uckel (DH) theory~\cite{DH}.

In this work, we investigate a different scenario and consider the linear response of a bulk electrolyte in a uniform flow.
Such a flow can correspond to an electrolyte flowing past a stationary object, {\it e.g.}, an optically trapped colloid or particle. Alternatively, we can think of a source moving in an otherwise stationary electrolyte. This source, for example, can be an active biomolecule or the tip of an atomic force microscope (AFM).

We present several generalized response functions of a system in a uniform and stationary flow, taking into account the combination of three different external sources: (i) a force density acting on the fluid, (ii) an externally imposed ionic chemical potential, and (iii) an external charge density.
Although the response of the velocity field is still independent of the electrostatic interactions, we show that the response of charge density, ionic number density, and their currents are coupled to the uniform flow. In particular, the response of the charge density to an external point charge is described by a nonlinear function of the fluid velocity, extending the DH  screening theory for a uniform flow.

The outline of this paper is as follows. In Sec.~\ref{sec2}, we present our model for a dilute electrolyte and a general external perturbation with a constant relative velocity. The basic electro-hydrodynamic
equations of our model are derived (Sec.~\ref{ssec2a}), and the linearized scheme is described (Sec.~\ref{ssec2b}).
Next, in Sec.~\ref{sec3}, the general response functions are derived and analyzed in detail. They include the hydrodynamic response (Sec.~\ref{ssec3a}), the number density
response (Sec.~\ref{ssec3b}), and the electric response (Sec.~\ref{ssec3c}).
In Sec.~\ref{sec4}, we summarize our results and relate our findings
to possible experiments.

%%%%%%%%%
\section{Model}
\label{sec2}

\subsection{Three sources and their responses}

Consider a dilute ionic solution, consisting of  1:1 monovalent cations and anions of total bulk concentration $n_{0}$, immersed in a continuum solvent of dielectric \ra{permittivity} $\varepsilon$. The solvent is modeled as an incompressible fluid with viscosity $\eta$. The ions are assumed to be point-like, and the friction coefficient of both cations and anions with the solvent is $\zeta$. The system is held at constant temperature $T$. Under these conditions, the homogeneous solution is in thermal equilibrium and satisfies local electroneutrality.

The homogeneous electrolyte can be perturbed by an externally controlled source.  When the object is at rest, the system reaches a new equilibrium state. However, when it is mobile \ra{at a constant velocity}, a steady state can be established, where the system is out of equilibrium, but all physical quantities are time-independent.  We explore this latter scenario of a mobile external perturbation with a relative velocity with respect to the ionic solution. For convenience, the frame of reference is chosen such that the perturbation is at rest, while the electrolyte flows with a velocity $\buo$.

\begin{figure*}[ht]
\centering
\includegraphics[width=0.85\textwidth]{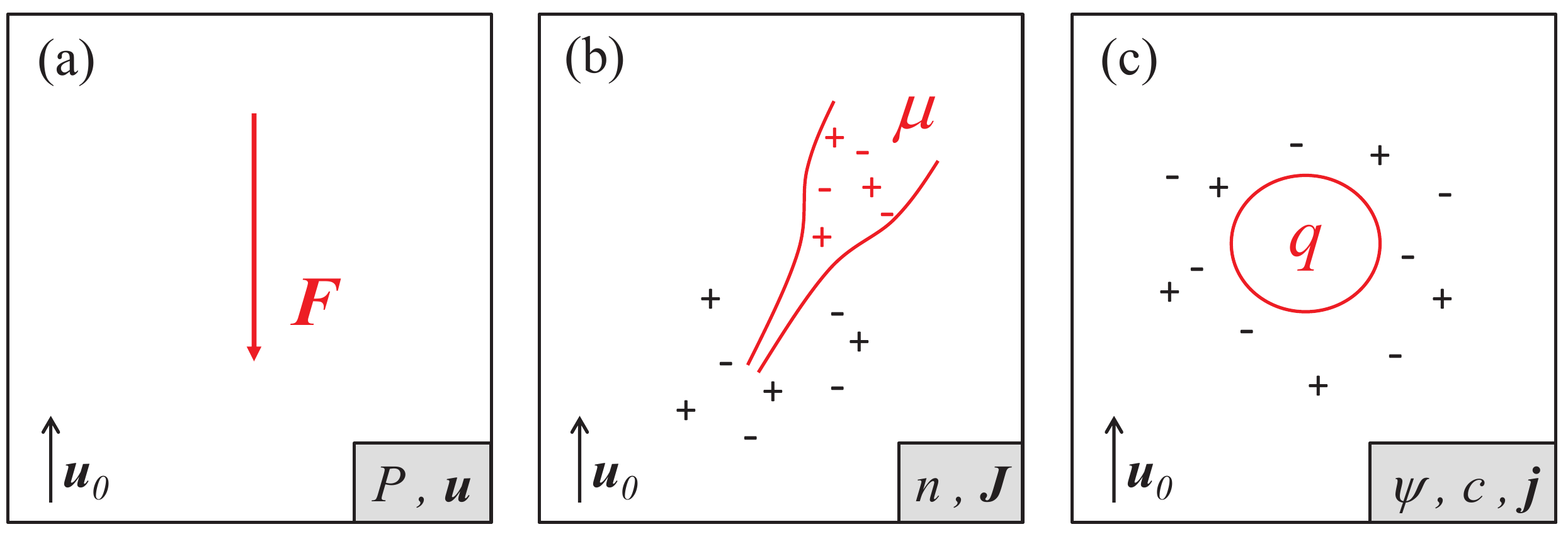}
\caption{(Color online) A schematic illustration of three possible external sources (in red) and their corresponding induced fields (in grey box). All the sources are at rest, while the electrolyte flows with a constant velocity $\buo=u_0\hat{\boldsymbol{z}}$. (a) A thin rod exerting a force density, $\bF$, in the negative $z$-direction, leading to a pressure, $P$, and velocity field, $\bu$, in the electrolyte.
(b) A tip of a pipette containing the electrolyte with a different ionic chemical potential, $\mu$, causing changes in the total ionic number density, $n$, and number density current, $\bjn$. (c) A colloid with charge density, $q$, producing an electrostatic potential, $\psi$, as well as a charge density, $c$, and electric current, $\bjc$. \ra{The partition into sources and consequent fields is described in detail in Sec.~\ref{ssec2b}.} Any surface effects stemming from the boundaries of the sources are neglected.
}
\label{fig1}
\end{figure*}

As a general perturbation, we consider the combination of three possible sources depicted in Fig.~\ref{fig1}: (a) an external force density, $\bF$, {\it e.g.}, a force exerted by a thin rod immersed in the solution, (b) an externally imposed ionic chemical potential, $\mu$, {\it e.g.}, a tip of a pipette containing the electrolyte solution with a different ionic concentration, and (c) an external charge density (per unit volume), $q$, {\it e.g.}, a small charged colloid.

We assume that surface effects from the sources are negligible, such that no boundary conditions are imposed. This assumption is appropriate for small sources or for the linear response in the far-field away from the sources. For example, a colloidal probe AFM with a fixed scanning velocity can be modeled as a combination of a point charge, $q$, and a point force density, $\bF$, originating from the drag on the colloidal probe.

Each of the three sources affects the electrolyte differently and can be identified with different fields that characterize the electrolyte, \ra{as is clarified in detail in Sec.~\ref{ssec2b} and} indicated in Fig.~\ref{fig1}. The force density in Fig.~\ref{fig1}(a), $\bF$, perturbs the electrolyte velocity, $\bu$, and pressure, $P$. In Fig.~\ref{fig1}(b), the externally imposed chemical potential, $\mu$, modifies the total ionic number density, $n$, and current, $\bjn$, defined as
%%%%%%%%%%%%%%
\begin{align}
\label{eq1}
n&=n_++n_- \, ,\nonumber\\
\bjn&=n_+\bv_+ +n_-\bv_- -n\bu \, .
\end{align}
%%%%%%%%%%%%%%
Here $n_\pm$ are the densities of cations and anions ($n_\pm=n_0$ in the homogeneous bulk), and $\bv_\pm$ are their velocities. Note that the current is evaluated in the moving frame of reference.

The external charge density of Fig.~\ref{fig1}(c), $q$, results in an electrostatic potential, $\psi$, a charge density (per unit volume), $c$, and an electric current, $\bjc$. The latter two are defined as
%%%%%%%%%%%%%%
\begin{align}
\label{eq2}
c&=e\left(n_+-n_-\right) \, ,\nonumber\\
\bjc&=e\left(n_+\bv_+ -n_-\bv_-\right) -c\bu \, ,
\end{align}
%%%%%%%%%%%%%%
where $e$ is the elementary charge. Similarly to the number density current $\bjn$, the electric current
$\bjc$ is defined according to the ionic relative velocities, $\bv_\pm-\bu$.
%%%%%%%%%%%%%fig1%%%%%%%%%%%%%
%%%%%%%%%%%%%%%%%%%%%%%%%%%%%%%%%

%%%%%%%%%%%%%%%%%%%%%%%%%
\subsection{Electro-hydrodynamic equations}
\label{ssec2a}

 The response of the seven fields, $P$, $\bu$, $n$, $\bjn$, $\psi$, $c$, and $\bjc$, to the three sources,
 $\bF$, $\mu$, and $q$, is obtained by solving seven coupled differential equations; the Stokes equation and incompressibility condition for the fluid, Poisson's equation for the electrostatic potential, force balance equations for the two ionic densities, and the corresponding continuity equations for the two currents. All of these equations can be derived consistently within a single framework, using Onsager's variational principle~\cite{Onsager1931,DoiBook}. Below, the equations are presented and discussed in detail.

First, the fluid is incompressible:
%%%%%%%%%%%%%
\begin{align}
\label{eq3}
\nabla \cdot \bu=0 \, .
\end{align}
%%%%%%%%%%%%%%%%%%%%%%%%%%%%%%%%%%
For low Reynolds numbers (no inertia), the electrolyte satisfies the Stokes equation,
%%%%%%%%%%%%%%%%%%%%%%%%%%%%%%%%%5
\begin{align}
\label{eq4}
\eta\nabla^{2}\boldsymbol{u}-\nabla P-
c\nabla \psi&
=-\boldsymbol{F} \, .
\end{align}
%%%%%%%%%%%%%%%%%%%%%%%%%%%%%%%%%%
The first term in Eq.~(\ref{eq4}) originates from the solvent viscosity, $\eta$. The third term stems from the solute charge, $c$, and couples the hydrodynamics with the electric variables.

It is possible to decompose the pressure $P$ and the force density $\bF$ into hydrodynamic and thermal terms. The pressure is given by the sum of the hydrodynamic solvent pressure and solute pressure as explained below. Similarly, the force density is the combined external force density on the electrolyte, $\boldsymbol{f}$, and diffusive thermal force, {\it i.e.}, $\boldsymbol{F}=\boldsymbol{f}-n\nabla \mu$. As mentioned above, $\mu$ is the externally imposed chemical potential of the ions, taken to be the same for the cations and anions.

The electrostatic potential, $\psi$, in Eq.~(\ref{eq4}) satisfies Poisson's equation (in SI units),
%%%%%%%%%%%%%%%%%%%%%%%%%%%%%%%%%%%%%
\begin{align}
\label{eq5}
\varepsilon\nabla^{2}\psi + c&=-q \, ,
\end{align}
%%%%%%%%%%%%%%%%%%%%%%%%%%%%%%%%%%%%%
where $\eps$ is the dielectric \ra{permittivity} of the medium and $q$ is the external charge density. Including both the contributions of the electric field, $\boldsymbol{E}=-\nabla \psi$,  and the hydrodynamic drag, we obtain
the force balance equations for the cations and anions as
%%%%%%%%%%%%%%%%%%%%%%%%%%%%%%%%%%%%%%
\begin{align}
\label{eq6}
- en_{+}\nabla \psi - \zeta n_{+}\left(\boldsymbol{v}_{+}-\boldsymbol{u}\right)-\kbt\nabla n_{+} & =n_{+}\nabla \mu \, ,\nonumber\\
e n_{-}\nabla \psi -
 \zeta n_{-}\left(\boldsymbol{v}_{-}-\boldsymbol{u}\right)-\kbt\nabla n_{-} & =n_{-}\nabla \mu \, ,
\end{align}
%%%%%%%%%%%%%%%%%%%%%%%%%%%%%%%%%%%%%%%%%%%
where $\kbt$ is the thermal energy. We combine the force balance equations above into two new equations in terms of the number density current, $\bjn$, and the electric current, $\bjc$:
%%%%%%%%%%%%
\begin{align}
\label{eq7}
\zeta\boldsymbol{J}+c\nabla \psi
+\kbt\nabla n&=-n\nabla\mu\, ,\nonumber \\
\zeta\boldsymbol{j}+e^2 n\nabla \psi
+\kbt\nabla c&=-c\nabla\mu \,.
\end{align}
%%%%%%%%%%%%%%
These equations couple the number density, $n$, and current, $\bjn$, with the electric variables, $\psi$, $c$, and $\bjc$ (electrostatic potential, charge density, and electric current, respectively).

Note that the term $\kbt\nabla  n_{\pm}$ in Eq.~(\ref{eq6}) that is carried over to Eq.~(\ref{eq7}), can be regarded
as the gradient of van t' Hoff ideal gas pressure. Here we assume that the system is well described by such a term, as it is not far from thermal equilibrium. It follows from the ideal-gas pressure form that steric effects and fluctuations of the electrostatic potential beyond the mean-field treatment are not included within our framework.

Finally, we combine the continuity equation for each of the two ionic flows, $\nabla \cdot\left(n_{\pm}\boldsymbol{v}_{\pm}\right)=0$, and arrive at the following equations in terms of the two currents:
%%%%%%%%%%%%%%%%%%%%%%%%%%%%%%%%%%%%%%
\begin{align}
\label{eq8}
\nabla \cdot\boldsymbol{J}+\bu\cdot\nabla n&=0 \, ,\nonumber \\
\nabla \cdot\boldsymbol{j}+\bu\cdot\nabla c&=0 \, .
\end{align}
%%%%%%%%%%%%%%%%%%%%%%%%%%%%%%%%%%%%%%
In principle, additional source terms can be included in the right-hand side of Eq.~(\ref{eq8})~\cite{Elrick1962,Okubo1969}. However, in this study, we neglect such extra ionic source terms, as well as any possible chemical reactions that lead to additional fluxes in the continuity equation.

In the absence of electrolyte flow and an externally imposed chemical potential, {\it i.e.}, $\bu=0$ and $\mu=0$, our set of equations reduces to the steady state Poisson-Nernst-Planck equations~\cite{Bazant04}. In thermodynamic equilibrium, these equations reduce further to the Poisson-Boltzmann equation~\cite{Bazant04,David95}. Finally, for small electrostatic potentials,
the Poisson-Boltzmann equation can be approximated by its linear form, which is the DH
equation~\cite{David95}.

It is convenient to rescale all variables and define the following dimensionless quantities:
%%%%%%%%%%%%%%
\begin{align}
 \label{eq9}
&\widetilde{P}\equiv\frac{P}{n_0\kbt}\, ,~\widetilde{\boldsymbol{u}}\equiv \frac{\zeta \ld }{\kbt}\boldsymbol{u} \, ,~\widetilde{\eta}\equiv \frac{4\pi\lb}{\zeta}\eta \, ,\nonumber\\
&\widetilde{n}\equiv \frac{n}{n_0}\, ,~\widetilde{\boldsymbol{J}}\equiv\frac{\zeta \ld}{n_0\kbt}\boldsymbol{J} \, ,~\widetilde{\mu}\equiv \frac{\mu}{\kbt} \, ,~\widetilde{\boldsymbol{F}}\equiv\frac{\ld}{n_0\kbt}\boldsymbol{F} \, ,\nonumber\\
&\widetilde{\psi} \equiv \frac{e\psi}{\kbt}\, ,~\widetilde{c}\equiv \frac{c}{en_0} \, ,~\widetilde{\boldsymbol{j}}\equiv\frac{\zeta \ld}{en_0\kbt}\boldsymbol{j} \, ,~\widetilde{q}\equiv \frac{q}{e n_0} \,,
 \end{align}
 %%%%%%%%%%%%%%%
where $\lb=e^2/\left(4\pi\eps\kbt\right)$ is the Bjerrum length and $\ld=(4\pi \lb n_0)^{-1/2}$ is the Debye screening length ($n_0$ is defined as the combined concentration of cations and anions together). Similarly, the position vector is rescaled as $\widetilde{\vecr}\equiv\vecr/\ld$ so that $\widetilde{\nabla}\equiv\ld\nabla$.

Hereafter, the tilde notation is omitted and all variables are treated as dimensionless quantities. Hence, the set of equations, Eqs.~(\ref{eq3})--(\ref{eq8}), can be written in a compact form as
%%%%%%%%%%%%
\begin{align}
\label{eq10}
\eta\nabla^{2}\boldsymbol{u}-\nabla P-c\nabla \psi&=
-\boldsymbol{F} \, ,\nonumber \\
\nabla^{2}\psi+c&=-q \, ,\nonumber \\ \boldsymbol{J}+c\nabla \psi
+\nabla n&=-n\nabla\mu\, ,\nonumber \\
\boldsymbol{j}+n\nabla \psi
+\nabla c&=-c\nabla\mu \, ,\nonumber \\
\nabla \cdot\boldsymbol{u}&=0 \, ,\nonumber \\
\nabla \cdot\boldsymbol{J}
+\bu\cdot\nabla n&=0 \, ,\nonumber \\
\nabla \cdot\boldsymbol{j}
+\bu\cdot\nabla c&=0 \, .
\end{align}
%%%%%%%%%%
 A general solution of these equations requires to specify the boundary conditions in terms of fixed (and charged) objects and interfaces. However, as we are interested only in bulk properties far from any boundaries, such conditions will not enter into our study.

%%%%%%%%%%%%%%%%%%%%%%%%%%%%%%%%%%%%%%%%%%
\subsection{Linearized equations}
\label{ssec2b}
%%%%%%%%%%%%%%%%%%%%%%%%%%%%%%%%%%%%%%%%%%%

In the absence of sources, $\boldsymbol{F}=q=\mu=0$, all the equations in Eq.~(\ref{eq10})  are homogeneous and describe a uniform bulk electrolyte in a constant flow. The ions are at rest in the moving fluid reference frame. They are distributed homogeneously and satisfy local electro-neutrality. This special solution is given by  $c=\psi=\boldsymbol{J}=\boldsymbol{j}=0$ while $\bu=\buo$ and $n=P=1$ (for the dimensionless variables).

We now analyze the system response to a small perturbation, as is described above, via a linearization of Eq.~(\ref{eq10}) around the homogeneous solution. Each field, $\Phi(\vecr)$, can be expanded as
$\Phi(\vecr)=\Phi_0+\Phi_1(\vecr)+\cdots$, where $\Phi_0$ is the homogeneous field and $\Phi_1(\vecr)$ is the linear correction in the presence of sources. All the linear corrections as well as the sources are considered to be of a similar small magnitude. We keep only terms that are linear in this small magnitude and neglect quadratic or higher-order terms. The linearization procedure is described in detail in Appendix~\ref{appA}.

The linearized equations can be written as three decoupled sets of equations. The first set includes the hydrodynamic variables [see Fig.~\ref{fig1}(a)] and reads
 %%%%%%%%%%%%%%
\begin{align}
\label{eq11}
\eta\nabla^2\bu_1(\vecr)-\nabla P_1(\vecr)&=
-\boldsymbol{F}(\vecr) \, ,\nonumber\\
\nabla \cdot \bu_1(\vecr)&=0 \, .
\end{align}
%%%%%%%%%%%%%%
These are the Stokes equation and incompressibility condition for the electrolyte. They describe how the force density, $\bF$, induces a pressure gradient that results in a fluid flow. These equations are independent of electrostatics, while the ionic number density (and not charge density) enters the definition of the pressure, $P$, and force density, $\bF$.

The second set of equations corresponds to the number density, $n_1$, and current, $\boldsymbol{J}_1$ [see Fig.~\ref{fig1}(b)]:
%%%%%%%%%%%%%%
\begin{align}
\label{eq12}
\bjn_1(\vecr)+\nabla n_1(\vecr)&=
-\nabla\mu(\vecr) \, ,\nonumber\\
\nabla\cdot\bjn_1(\vecr)+\buo\cdot\nabla n_1(\vecr)&=0 \, .
\end{align}
%%%%%%%%%%%%%%
The first equation describes how the externally imposed chemical potential, $\mu$, induces a number density gradient, $\nabla n_1$, and consequently yields a number density current, $\bjn_1$. The second equation in Eq.~(\ref{eq12}) implies that the current $\bjn_1$ is a compressible field, since it is defined relative to the fluid velocity. \raa{Combining the two equations yields a steady-state convection-diffusion equation~\cite{Masliyah} for the number density, $n_1$.}

Finally, we obtain for the electric variables [see Fig.~\ref{fig1}(c)]
%%%%%%%%%%%%%%
\begin{align}
\label{eq13}
\nabla^2\psi_1(\vecr)+c_1(\vecr)&=-q(\vecr) \, ,\nonumber\\
\bjc_1(\vecr)+\nabla c_1(\vecr)&=-\nabla\psi_1(\vecr) \, ,\nonumber\\
\nabla\cdot\bjc_1(\vecr)+\buo\cdot\nabla c_1(\vecr)&=0 \, .
\end{align}
%%%%%%%%%%%%%%
The three equations above describe how an external charge, $q$, produces an electric field and a charge density gradient, whose combined effect leads to an electric current. The first equation in Eq.~(\ref{eq13}) is Poisson's equation that remains unchanged by the linearization procedure. The bottom two equations have the same structure as in Eq.~(\ref{eq12}), where $\psi_1$ plays the role of $\mu$. The reason is that the chemical potential and number density are conjugate variables, just as the charge density and electrostatic potential are.

%%%%%%%%%%%%%%%%%%%%%%%%%%%%%%%%%%%
\section{Linear Response Functions}
\label{sec3}
%%%%%%%%%%%%%%%%%%%%%%%%%%%%%%%%

The solutions to each of the three sets of linear equations in Eqs.~(\ref{eq11})--(\ref{eq13}) can be written in terms of linear response functions. Such solutions are conveniently found via the Fourier transform of the fields and sources, where the Fourier transform of any function $\Phi(\vecr)$ is defined as $\Phi(\veck)=\int\D^3r\,\Phi(\vecr)\exp(-i\veck{\cdot}\vecr)$.
In Fourier space, the response of the system in terms of a variable $\Phi$ to an external source $S_\Phi$ is given by the product, $\Phi(\veck)=\mathcal{K}_\Phi(\veck)S_\Phi(\veck)$, where $\mathcal{K}_\Phi(\veck)$ is the response
function (kernel). In real space, the same response has a convolution form:
%%%%%%%%%%%%%%%%%
\begin{align}
\label{eq14}
\Phi(\vecr)=\int \D^3r' \,\mathcal{K}_\Phi\left(\vecr-\vecr' \right)S_\Phi(\vecr') \, .
\end{align}
%%%%%%%%%%%%%%%

Below we present separately the response functions for each of the three sets: (i) the response of the hydrodynamic variables, $\bu_1$ and $P_1$, to the source $\bF$, (ii) the response of the number density and current variables, $n_1$ and $\bjn_1$, to the source $\mu$, and (iii) the response of electric variables, $\psi_1$, $c_1$ and $\bjc_1$, to the source $q$. For a full derivation of these response functions, see  Appendix~\ref{appA}.

%%%%%%%%%%%%%%%%%%%%%%%%%%%
\subsection{Response of hydrodynamic variables}
\label{ssec3a}
%%%%%%%%%%%%%%%%%%%%%%%%%%%

%%%%%%%%%%%%%%%fig2%%%%%%%%%%%%%
\begin{figure*}[ht]
\centering
\includegraphics[width=0.85\textwidth]{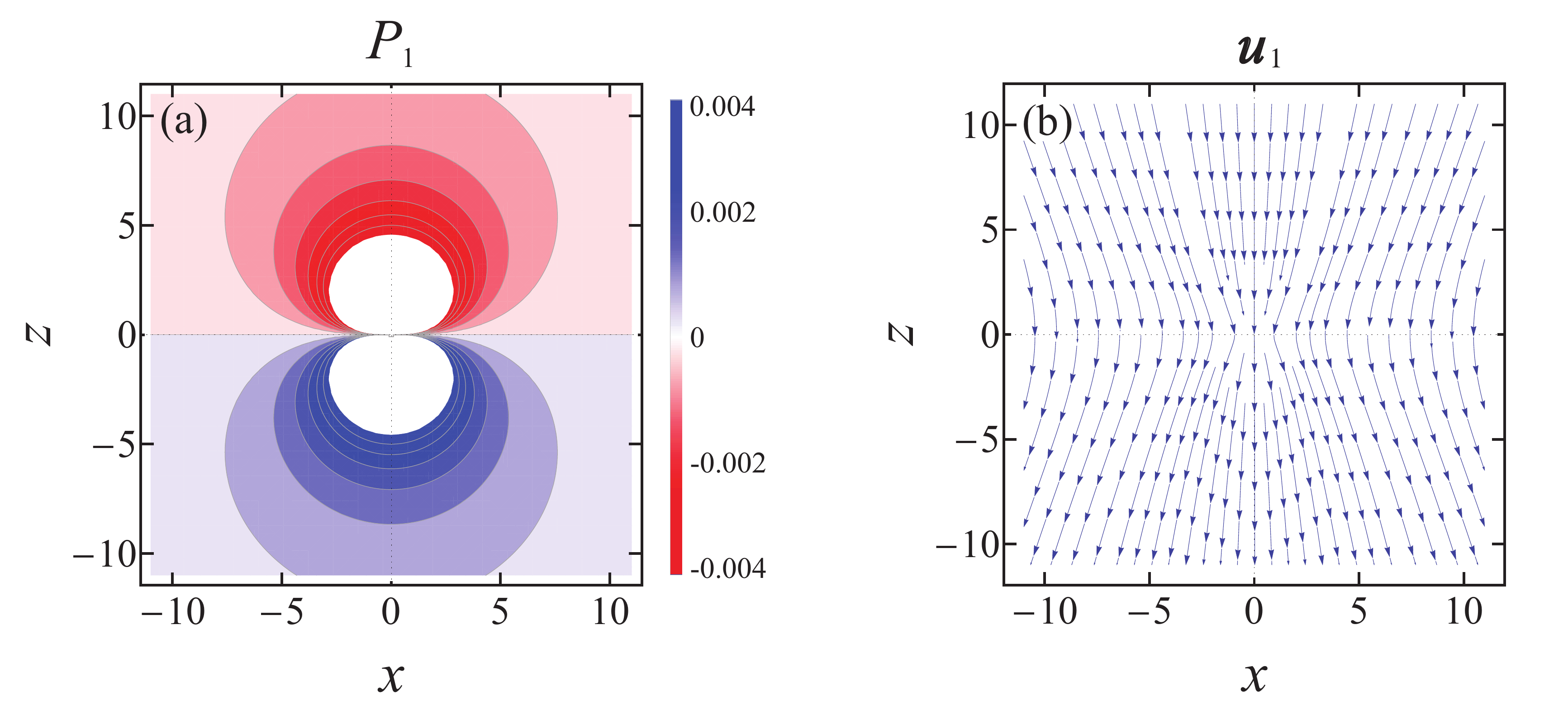}
\caption{(Color online) Response of the hydrodynamic fields, pressure $P_1$ and fluid velocity $\bu_1$, to a point force source at the origin, $\bF=-\delta(\vecr)\hat{\bz}$ [see Eq.~(\ref{eq16})]. All quantities are dimensionless, according to the definitions of Eq.~(\ref{eq9}). The response is plotted in the $xz$-plane, and can be extended to the entire space due to the azimuthal symmetry around the $z$-axis.  (a) Contour plots of equal pressure. The red regions in the upper half-plane correspond to negative values, $P_1<0$, while the blue regions in the bottom half-plane correspond to positive ones, $P_1>0$. Note that the inner area corresponds to larger absolute values of $P_1$ and is left white for clarity sake. (b) Stream lines of the velocity $\bu_1$. The arrows indicate the flow direction.  }
\label{fig2}
\end{figure*}
%%%%%%%%%%%%%%%%%%%%%%%%%%%%%%%%%
Consider an external force density imposed on the fluid, $\bF$, as in Fig.~\ref{fig1}(a). Solving Eq.~(\ref{eq11}) for the hydrodynamic response (Appendix~\ref{appA}), we arrive at
%%%%%%%%%%%%%%%%%%%%%%%%
\begin{align}
\label{eq15}
P_1(\veck)&=-\frac{i}{k^2}\veck\cdot\bF(\veck) \, ,\nonumber\\
\bu_1(\veck)&=\frac{1}{\eta k^2}\left(\mathbf{I}
-\frac{\veck\veck}{k^2}\right) \cdot \bF(\veck) \, ,
\end{align}
%%%%%%%%%%%%%%%%%%%%%%%%
where $\mathbf{I}$ is the identity tensor, $k=\vert \veck \vert$, and $\veck\veck$ is the dyadic product ($2$nd rank tensor). Note that Eq.~(\ref{eq15}) is independent of ionic properties.

Transforming the hydrodynamic responses of Eq.~(\ref{eq15}) into real space yields the response functions
%%%%%%%%%%%%%%%%%%%%%%%%%%%%%%
\begin{align}
\label{eq16}
\boldsymbol{\mathcal{K}}_P(\vecr)&= \frac{\vecr}{4\pi r^3}  \, ,\nonumber\\
\boldsymbol{\mathcal{K}}_{\bu}(\vecr)&=
\frac{1}{8\pi\eta r}\left(\mathbf{I}+\frac{\vecr \vecr}{r^2}\right) \, ,
\end{align}
%%%%%%%%%%%%%%%%%%%%%%%%%%%%%%5
where $r=\vert \vecr \vert$. We emphasize that $\boldsymbol{\mathcal{K}}_P$ is a vector and $\boldsymbol{\mathcal{K}}_{\bu}$ is a $2$nd rank tensor.  As the linearized hydrodynamic equations are independent of electrostatics, Eq.~(\ref{eq16}) restores the Oseen's result with $\boldsymbol{\mathcal{K}}_{\bu}$ being the Oseen tensor~\cite{Oseen,Happel}. The response to a point force source $\bF=-\delta(\vecr)\hat{\bz}$
($\boldsymbol{\hat{z}}$ is a unit vector along the $z$-direction)
of Eq.~(\ref{eq16}) is illustrated in Fig.~\ref{fig2}.
As the Oseen's result is well-known, we present it only for completeness and do not discuss it further.
%We rather focus on the number density and current response as well as the response of electric variables, as is shown below.

%%%%%%%%%%%%%%%%%%%%%%%%%%%%%%%%%%%%%%
\subsection{Response of number density and current}
\label{ssec3b}
%%%%%%%%%%%%%%%fig3%%%%%%%%%%%%%

%%%%%%%%%%%%%%%%%%%%%%%%%%%%%%%%%%%%%%%%%%%%%%%%%%%%%%%%%%%%%
\begin{figure*}[ht]
\centering
\includegraphics[width=0.85\textwidth]{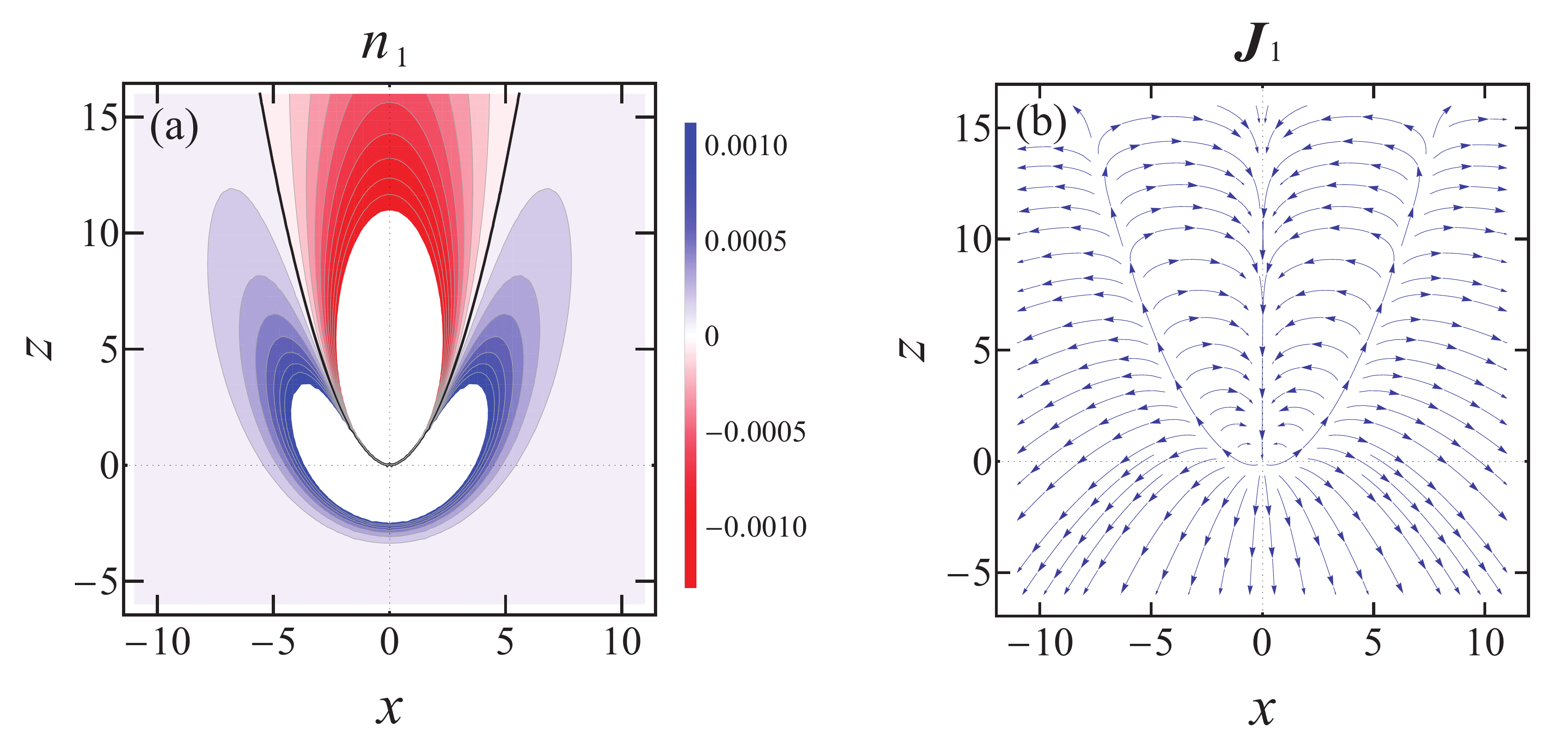}
\caption{(Color online) Response of the number density, $n_1$, and current field, $\bjn_1$, to a point source at the origin, $\mu=\delta(\vecr)$,  for the fluid velocity $\buo=2\hat{\bz}$ [see Eqs.~(\ref{eq18}) and (\ref{eq19})].  (a) Contour plots of equal number densities $n_1$. The contour $n_1=0$ is drawn by the black line [see Eq.~(\ref{eq20})]. Note that the inner (uncolored) areas correspond to even larger absolute values of $n_1$. (b) Stream lines of the number density current, $\bjn_1$.}
\label{fig3}
\end{figure*}
%%%%%%%%%%%%%%%%%%%%%%%%%%%%%%%%%

Consider an externally imposed chemical potential, $\mu$, of the ionic species, as in Fig.~\ref{fig1}(b). The response to such a source is given  by solving Eq.~(\ref{eq12}) (see also Appendix~\ref{appA}),
%%%%%%%%
\begin{align}
\label{eq17}
n_1(\veck)&=-\frac{k^2}{k^2+i\buo\cdot\veck}\,\mu(\veck) \, ,\nonumber\\
\bjn_1(\veck)&=\frac{\left(\buo\cdot\veck\right)\veck}{k^2+i\buo\cdot\veck} \,\mu(\veck) \, .
\end{align}
%%%%%%%%%%%
We see here that the two fields are related by $\bjn_1(\veck)=-(\buo\cdot\hat{\veck})\hat{\veck}\,n_1(\veck)$, where $\hat{\veck}=\veck/k$ is a unit vector.

In the absence of flow, $\buo=0$, the response reduces to $n_1=-\mu$ and $\bjn_1=0$.
Such a response can be obtained from the equilibrium Boltzmann distribution for small $\mu$ values because $n_1={\rm e}^{-\mu}-1\approx-\mu$. Otherwise, for a non-zero velocity, the direction of $\buo$ (taken in the $z$-direction) defines an axis of symmetry, and the usual spherical symmetry becomes an azimuthal one (body of revolution around $\buo$).

The real-space response of the number density, $n_1$, and current, $\bjn_1$, are given by the two
response functions
%%%%%%%%%%%%%%%%%%%%
\begin{align}
\label{eq18}
\KC_n(\vecr)&=-\delta(\vecr)+g(\vecr) \, ,\nonumber\\
\boldsymbol{\KC}_{\boldsymbol{J}}(\vecr)&=-\nabla g(\vecr) \, ,
\end{align}
%%%%%%%%%%%%%%%%%%%%%
where $\KC_n$ is a scalar, $\boldsymbol{\KC}_{\boldsymbol{J}}$ is a vector, and the function $g$ is given by
%%%%%%%%%%%%%%%%%%%%%%%
\begin{align}
\label{eq19}
g(\vecr)=u_0\frac{u_0 r^2-z(2+u_0 r)}{8\pi r^3}\e^{u_0(z-r)/2} \, .
\end{align}
%%%%%%%%%%%%%%%%%%%%
\raa{We note that the Green's function for the corresponding two-dimensional convection-diffusion equation was solved in Ref.~\cite{Choi05}.}

The expressions of Eqs.~(\ref{eq18}) and (\ref{eq19}) for $n$ and $\bjn$ are plotted in Fig.~\ref{fig3}. It is convenient to interpret this case by considering a source moving in an otherwise stationary electrolyte with a velocity $-\buo$.
Due to friction, ions are pushed by the source outside of its trajectory into an outer region. The boundary of this region can be obtained from Eq.~(\ref{eq19}) as the contour of $n_1=0$ and is given by
%%%%%%%%%%%%%%%%%
\begin{align}
\label{eq20}
r=\frac{2}{u_0}\frac{\cos\theta}{1-\cos\theta} \, .
\end{align}
%%%%%%%%%%%%%%%%
In the above, $\theta$ is the azimuthal angle between $\vecr$ and $\hat{\boldsymbol{z}}$, and this contour is drawn as a black line in Fig.~\ref{fig3}(a). Consequently, a wake is formed, as is seen in Fig.~\ref{fig3}(b). Ions within the wake experience a rotational flow, while ions in the outer region flow away from the moving source, with an average velocity in the negative $z$-direction.

%%%%%%%%%%%%%%%%%%%%%%%%%%%%%%%%%%%%%%%
\subsection{Response of electric variables}
\label{ssec3c}
%%%%%%%%%%%%%%%%%%%%%%%%%%%%%%%%

%%%%%%%%%%%%%%%fig5%%%%%%%%%%%%%
\begin{figure*}[ht]
\centering
\includegraphics[width=0.85\textwidth]{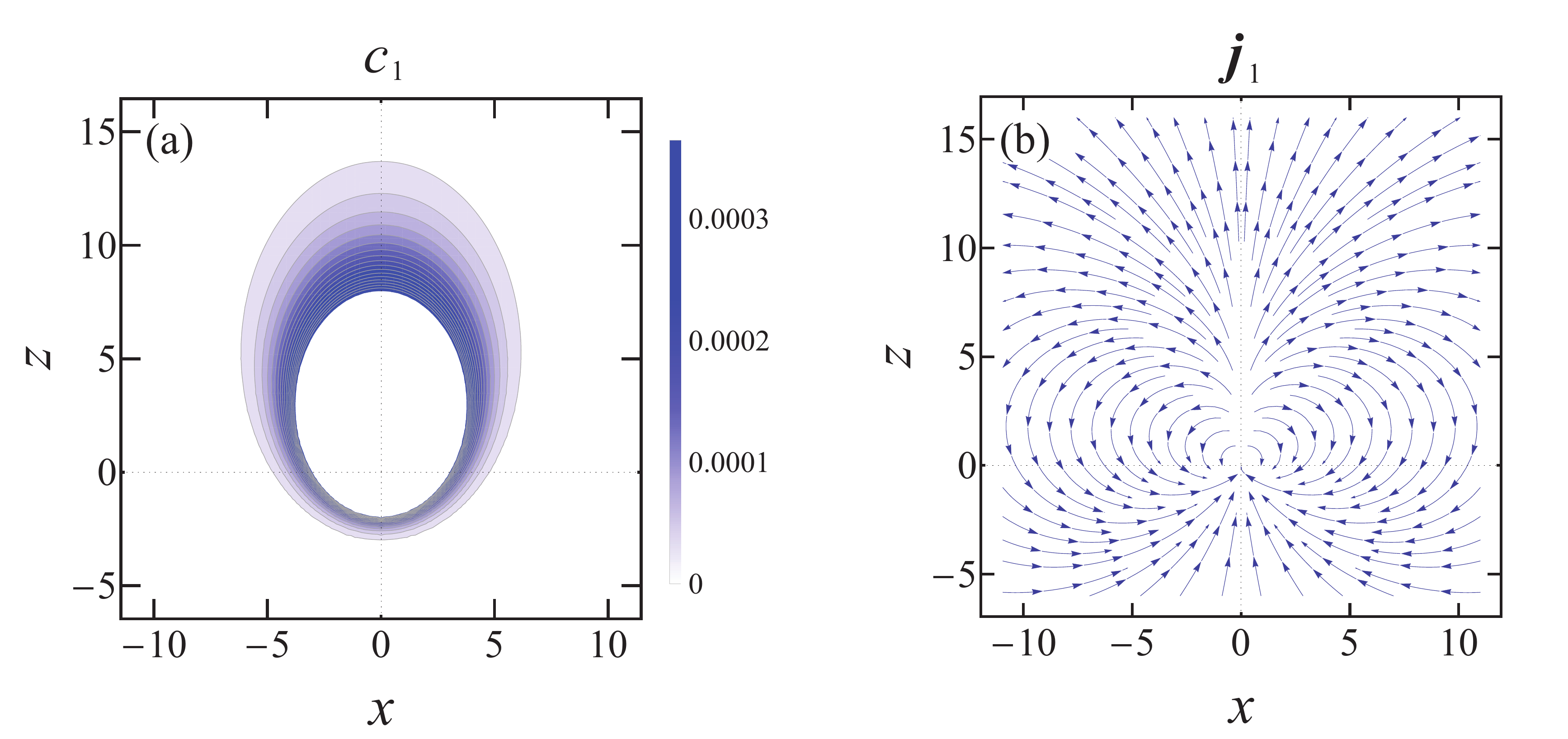}
\caption{(Color online) Response of the charge density, $c_1$, and electric current,  $\bjc_1$, to a negative point charge source at the origin, $q=-\delta(\vecr)$ for the fluid velocity $\buo=2\hat{\bz}$ [see Eq.~(\ref{eq22})]. (a) Contour plots of equal charge densities $c_1$. Note that the inner (uncolored) area corresponds to even larger absolute values of $c_1$.
(b) Stream lines of the electric current $\bjc_1$.}
\label{fig4}
\end{figure*}
%%%%%%%%%%%%%%%%%%%%%%%%%%%%%%%%%
We now consider the third case where the electrolyte is perturbed by an external charge source, $q$, as in Fig.~\ref{fig1}(c). In response, mobile ions surround the source and an electric current is formed. The charge density and electric current
obtained by solving Eq.~(\ref{eq13}) are given by
%%%%%%%%%%%%%%%%%%%%%%%%
\begin{align}
\label{eq21}
c_1(\veck)&= -\frac{1}{1+i\buo\cdot\veck+k^2}\,q(\veck) \, ,\nonumber\\
\bjc_1(\veck)&= \frac{(\buo\cdot\hat{\veck})\hat{\veck}}{1+i\buo\cdot\veck+k^2}\,q(\veck) \, .
\end{align}
%%%%%%%%%%%%%%%%%%%%%%%%
For $\veck=0$, Eq.~(\ref{eq21}) yields $c_1(\veck = 0)=-q(\veck=0)$, satisfying electro-neutrality.
In the limit of $\buo = 0$, on the other hand, the standard DH result is obtained, {\it i.e.}, $c_1(\veck)=-q(\veck)/(1+k^2)$.
The charge density and electric current of Eq.~(\ref{eq21}) are related to each other by $\bjc_1(\veck)=-(\buo\cdot\hat{\veck})\hat{\veck}\,c_1(\veck)$, just as the number density and current in Eq.~(\ref{eq17})
are related.

Performing the inverse Fourier transform of Eq.~(\ref{eq21}) results in the real-space response functions
%%%%%%%%%%%%
\begin{align}
\label{eq22}
\KC_c(\vecr)&=-\frac{1}{4\pi r}
\exp\left(\frac{u_0 z}{2} -\sqrt{1+\frac{u_0^2}{4}}\,r\right) \, ,\nonumber\\
\boldsymbol{\mathcal{K}}_{\boldsymbol{j}}(\vecr)&=
\nabla\left[\buo\cdot\nabla\KC_c(\vecr)\right] \, ,
\end{align}
%%%%%%%%%%%%
where $\KC_c$ is a scalar and $\boldsymbol{\mathcal{K}}_{\boldsymbol{j}}$ is a vector.
The first line of Eq.~(\ref{eq22}) describes the ``relaxation effect" of the ionic cloud. Ions are dragged in the direction of the flow, and the otherwise spherical charge density of mobile ions is stretched in the flow direction, as is plotted in Fig.~\ref{fig4}(a). This effect is known as the Dorn effect~\cite{Booth} in the case of sedimentation, where counterions accumulate behind the sedimenting particle.  A similar distortion of the ionic cloud has been previously reported in the presence of a shear flow~\cite{Lever1979}.

As a consequence of the relaxation effect, the charged source together with the ionic cloud have a non zero dipole moment. Assuming a point-like source $q(\vecr)=Q\delta(\vecr)$, we integrate $\KC_c(\vecr)\vecr$ and find the dipole moment $-Q\bu_0$. \ra{Note that all the physical quantities here are dimensionless, as is explained above.} For example, a negatively charged source moving in the negative $z$-direction results in a dipolar moment in the positive $z$-direction, as can be inferred from Fig.~\ref{fig4}(a).
At large distances, this dipole dominates the electric field, as is further discussed below.
%The dipolar moment can also be obtained by considering the leading-order term in $k$ of Eq.~(\ref{eq21}), which is $c_1(\veck)\approx i\left(\buo\cdot\veck\right) q(\veck)$. This expression transforms into $\buo\cdot\nabla q(\vecr)$ in real space and corresponds to the same dipole moment, $-q\bu_0$.

While  the ionic cloud is well approximated by a dipole far from the sources, the exact form of the charge density becomes important closer to it. By examining the argument of the exponent in Eq.~(\ref{eq22}), it is possible to approximate the shape of the ionic cloud by a prolate ellipsoid, satisfying the equation
%%%%%%%%%%%%%
\begin{align}
\label{eq23}
\frac{\rho^2}{a^2}+\frac{1}{a^2\left(1+u_0^2/4\right)}\left(z-\frac{u_0a}{2}\right)^2=1 \, .
\end{align}
%%%%%%%%%%%%
In the above equation, $\rho=\sqrt{x^2+y^2}$ is the absolute value of the in-plane two-dimensional vector $\boldsymbol{\rho}=\left(x,y\right)$, while $a$  is the minor semi-axis and $a \sqrt{1+u_0^2/4}$ is the major one. Both the major axis and one of the foci of the ellipsoid increase with $u_0$. Such a description fits well with the ionic cloud illustrated in Fig.~\ref{fig4}(a).

Another important feature of the non-equilibrium steady state is the formation of an electric current, as is given by the second line of Eq.~(\ref{eq22}). An example for such a current is plotted in Fig.~\ref{fig4}(b). The symmetric closed loops around the $z$-axis resemble those of a dipolar field. This observation can be better understood by examining Eq.~(\ref{eq21}) for large distances. Keeping only the lowest-order of $k$ leads to the electric current response, $(\buo\cdot\hat{\veck})\hat{\veck}$. The inverse Fourier transform of this function is obtained as a second derivative of the kernel
%%%%%%%%%%%%%%
\begin{align}
\label{eq24}
\boldsymbol{\mathcal{K}}_{\boldsymbol{j}}(\vecr)\approx \frac{1}{4\pi r^3}
\left[ \buo-3\left(\buo\cdot\hat{\vecr}\right)\hat{\vecr} \right]\, ,
\end{align}
%%%%%%%%%%
which indeed corresponds to a field of a dipolar moment $-\buo$.

%%%%%%%%%%%%%fig5%%%%%%%%%%%%%
\begin{figure*}[ht]
\centering
\includegraphics[width=0.85\textwidth]{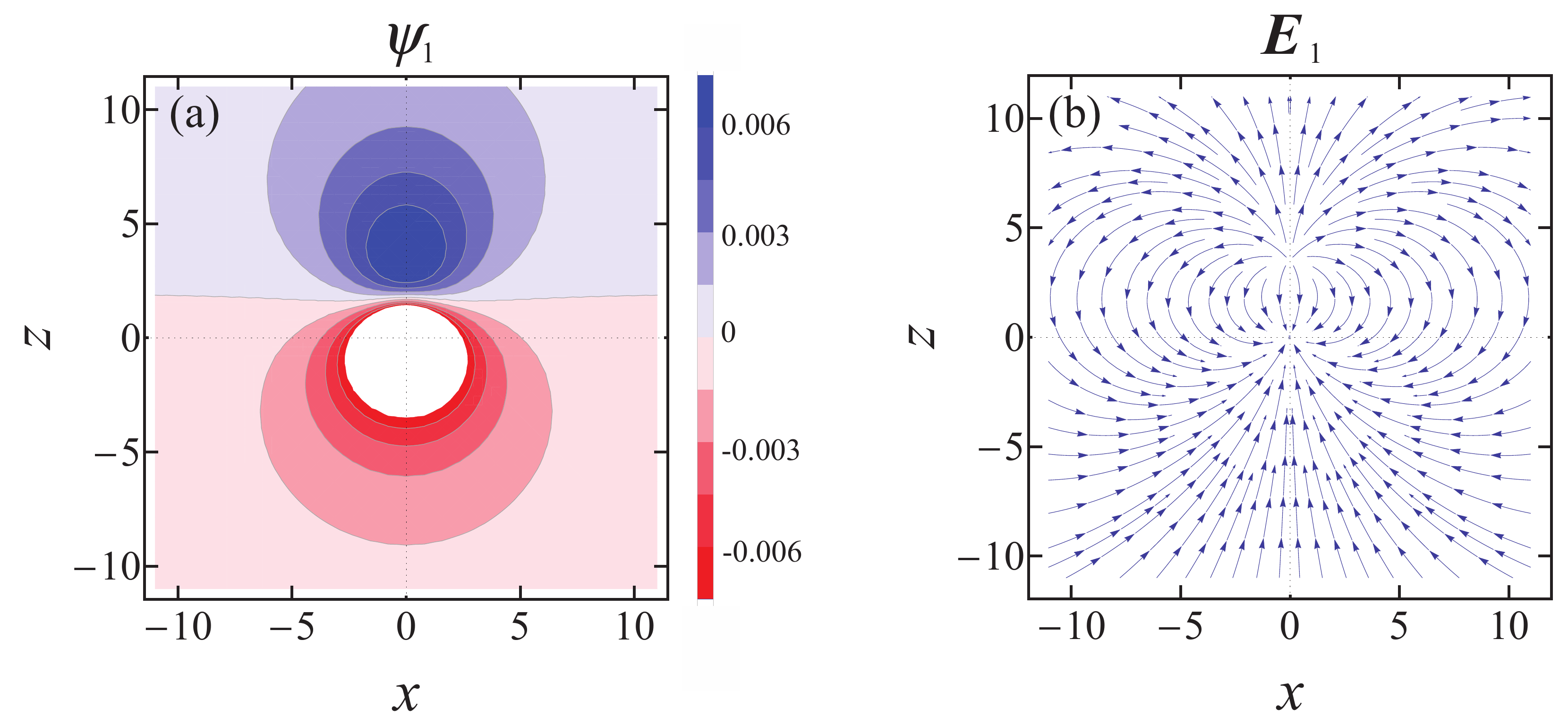}
\caption{(Color online) Response of  of the electrostatic potential, $\psi_1$, and electric field, $\boldsymbol{E}_1$, to a negative point charge source at the origin $q=-\delta(\vecr)$ for the fluid velocity $\buo=2\hat{\bz}$ [see
Eqs.~(\ref{eq27}) and (\ref{eq28})].   (a) Contour plots of equipotential values of $\psi_1$.
Note that the inner (uncolored) area corresponds to even larger absolute values of $\psi_1$. (b) Stream lines of the electric field $\boldsymbol{E}_1$.
}
\label{fig5}
\end{figure*}
%%%%%%%%%%%%%%%%%%%%%%%%%%%%%%%%%
The charge density and electric current of Eqs.~(\ref{eq21}) and (\ref{eq22}) are accompanied by an electrostatic potential, $\psi_1$. By solving Eq.~(\ref{eq13}), we find that
%%%%%%%%%%%
\begin{align}
\label{eq25}
\psi_1(\veck)=\frac{1}{k^2}
\left(1-\frac{1}{1+i\buo\cdot\veck+k^2}\right)q(\veck) \, .
\end{align}
%%%%%%%%%%
The prefactor $1/k^{2}$ is the Fourier transform of the Coulomb kernel, $1/(4\pi r)$. The first term inside the parenthesis corresponds to the direct effect that a charged source has on the potential. As is evident from Eq.~(\ref{eq21}), the second term corresponds to the ionic cloud.
Furthermore, substituting $\psi_1(\veck)$ in Eq.~(\ref{eq25}) into Eq.~(\ref{eq21}) yields
%Furthermore, we recall from the differential equations for the number density and current [see Eq.~(\ref{eq12})] and for the electric variables [see Eq.~(\ref{eq13})] that $\psi_1$ acts as a source for $c_1$ and $\bjc_1$ in the same way that $\mu$ acts as a source for $n_1$ and $\bjn_1$. Indeed, substituting $\psi_1(\veck)$ in Eq.~(\ref{eq25}) into Eq.~(\ref{eq21}) yields
%%%%%%%%%%%
\begin{align}
\label{eq26}
c_1(\veck)&=-\frac{k^2}{k^2+i\buo\cdot\veck}\,
\psi_1(\veck) \, ,\nonumber\\
\bjc_1(\veck)&=\frac{(\buo\cdot\veck)\veck}
{k^2+i\buo\cdot\veck}\, \psi_1(\veck) \, ,
\end{align}
%%%%%%%%%%
demonstrating the same dependence on $\psi_1$ as that of $n_1$ and $\bjn_1$ on $\mu$ [see Eq.~(\ref{eq17})],
as mentioned before.

Equation~(\ref{eq25}) generalizes the DH result for sources moving with a relative velocity with respect to the solvent. We transform it back to real
space to obtain
%%%%%%%%%%%%%%%%
\begin{align}
\label{eq27}
\KC_\psi\left(\rho,z\right)&=\frac{1}{4\pi \sqrt{\rho^2+z^2}}\nonumber\\
&-\frac{1}{4\pi u_0}\int_{0}^{\infty}\D z'\,
\frac{1-\exp[h(\rho,z-z')]}{\sqrt{\rho^2+(z-z')^2}}\, \e^{-z'/u_0} \, ,
\end{align}
%%%%%%%%%%%%%%
where we have denoted
%%%%%%%%%%%%%%
\begin{align}
\label{eq28}
h(\rho,z)=\frac{u_0}{2}z-\sqrt{\left(1+\frac{u_0^2}{4}\right)
\left(\rho^2+z^2\right)} \, .
\end{align}
%%%%%%%%%%%%%%%%%%%%
The relaxation effect, therefore, carries over from the charge density to the electrostatic potential. The steady flow breaks the spherical symmetry between the $z$-direction and the $xy$-plane, leading to an over-screened potential in the positive $z$-direction and an under-screened one in the negative direction. The potential is also over-screened in the normal plane, as indicated above by the factor $(1+u_0^2/4)^{1/2}$.

The equipotential contours of $\psi_1$ are plotted in Fig.~\ref{fig5}(a), while the field lines of the resulting electric field, $\boldsymbol{E}_1=-\nabla\psi_1$, are plotted in Fig.~\ref{fig5}(b). The figure highlights the distinct features of the steady state solution both in the near and far fields. In the near field, the equipotential contours of $\psi_1$ resemble the ellipsoidal charge density of Fig.~\ref{fig4}(a), and describe an anisotropic screened interaction.
The far-field behavior, on the other hand, can be described by the dipole moment of the charge density, as discussed before.
Therefore, the electrostatic interaction is long-ranged and not screened as in the DH equilibrium case.

%Keeping the leading-order term in $k$ in Eq.~(\ref{eq25}) yields $\psi_1(\veck)\approx i\left(\buo\cdot\veck\right) q(\veck)/k^2$. For a point source, $q(\vecr)=q\delta(\vecr)$, this expression transforms into
%%%%%%%%%%%%%%
%\begin{align}
%\label{eq29}
%\psi_1(\vecr)\approx-\frac{q(\buo\cdot\vecr)}{4\pi r^3} \, ,
%\end{align}
%%%%%%%%%%%%%%
%which corresponds to an electric dipole moment of $-q\buo$, in accordance with our calculations for the charge density, $c_1$.
%For a colloidal suspension under a gravitational field, the collective contribution of such dipoles results in the sedimentation potential.

%%%%%%%%%%%%%%%%%%%%%%%%%%%%%%%%%%%%%%%%%%%%
\section{Summary and discussion}
\label{sec4}
%%%%%%%%%%%%%%%%%%%%%%%%%%%%%%%%%%%%%%
In this paper, we have presented a general formalism to analyze the steady-state linear response of bulk electrolyte solutions to an externally imposed source.
Our focus was on the effect of constant relative velocity between the electrolyte and source. In the absence of such a velocity, the derived response functions reduce to the Oseen's result for the hydrodynamic variables and the DH one for the electrostatic variables.

The steady flow is shown to break the natural spherical symmetry of the bulk and to define an azimuthal symmetry around the direction of motion. Namely, the hydrodynamic drag in presence of flow distorts the ionic cloud from a sphere to an ellipsoidal shape (relaxation effect) and consequently modifies the electric field close to the source and far from it.

Close to the source, the elongated ionic cloud exerts a net electric force in the direction of $\buo$.
If we consider a source of total charge $Q$ moving with velocity $-\buo$, the electric force thus opposes the direction of motion. For small velocities, this force scales as $\sim Q^2u_0$. This result can be obtained by expanding the electrostatic potential of Eq.~(\ref{eq27}) to linear order in $u_0$, and integrating the electric field over the surface of a small sphere around the origin. Such a force can be interpreted as an increase in the source friction coefficient. More generally, the change in viscosity of charged suspensions is known as the electroviscous effect~\cite{Ohshima}.

Far from the source, the anisotropic charge density together with the charged source can be described by an electric dipole moment $-Q\bu_0$, where $Q$ is the total external charge. Therefore far-field behavior of the electric field $\boldsymbol{E}_1$ and electric current $\bjc_1$ are well approximated by dipolar fields. Such long-ranged fields qualitatively differ from the equilibrium ones, where the electric field decays exponentially and no current arises.

The above effect is expected to be pronounced especially for large $u_0$ values. \raa{This dimensionless parameter quantifies the role of convection over diffusion, and can be interpreted as the P\'eclet number~\cite{Masliyah,Choi05}, with the Debye length, $\ld$, playing the role of the characteristic length scale in the system. The interplay between convection and diffusion is discussed further below. For the sake of clarity,}  we return to physical quantities in their natural units. In particular, we refer to $u_0$ below in units of velocity.

According to Eq.~(\ref{eq9}), high velocities correspond to $u_0>\kbt/(\zeta\ld)=D/\ld$, where $D$ is the diffusion coefficient. When this condition is satisfied, the thermal energy of ions is not sufficient to overcome the hydrodynamic drag across the Debye length, $\ld$, and ions cannot form a screening ionic cloud. As an alternative interpretation, high velocity values refer to $u_0>\ld/\tau$, where we have introduced the relaxation time (also called ``ambipolar time"), $\tau=\ld^2/D$, as the typical time for ions to diffuse throughout the Debye length~\cite{Bazant04}. Therefore, at high velocities, ions move too rapidly to screen a charged object properly.

We estimate the velocity scale as mentioned above, by considering a dilute aqueous solution at room temperature with $n_0=2$\,mM. The corresponding Debye length is $\ld\simeq10$\,nm. The diffusion coefficient of simple ions in such a solution is $D\approx10^{-9}$\,m$^2/$s~\cite{Vanysek}, leading to $\tau\simeq100$\,ns.  Then the ratio $\ld/\tau$ yields a velocity of $u_0\simeq0.1$\,m/s, which is relatively large. However, as $\ld/\tau\sim1/(\sqrt{\eps}\zeta)$, the velocity becomes much smaller for solvents that are more viscous than water even if they are less polar.

As an example for such a viscous solvent, we mention glycerol with permittivity $\eps\simeq40 \eps_0$ (where $\eps_0$ is the vacuum permittivity) and viscosity $\eta\simeq1.4$\,Pa$\cdot$s at $T=293$\,K~\cite{Glycerol}.  For the same ionic concentration of the aqueous solution above, the velocity $u_0\simeq100$\,$\mu$m/s is obtained. Velocities of order $100$\,$\mu$m/s are accessible in experiments and have been used in colloidal-probe AFM measurements~\cite{Lai,Darwiche,Honig} as the relative velocity between probe and object.

We note that the relaxation effect \ra{within our framework is continuous in $u_0$, without any critical behavior.} Furthermore, it is not expected to play a role in standard experiments designed to measure equilibrium forces. In these setups, it is customary to measure forces over a range of velocities, in order to ensure that the force is velocity-independent~\cite{Lhermerout}. In this manner,  possible electro-hydrodynamic effects are surpassed.

As a possible experimental setup to capture the dipolar electric interaction, we propose to measure  the relative translation between two identical, spherical and charged colloids, undergoing sedimentation in a large container. The hydrodynamic forces between two identical spheres in an unbound fluid induce no relative translation~\cite{Happel,Goldfirend16}, leaving the mutual electric force as the main possible origin for such a motion. \ra{For micron sized colloids sedimenting in a dilute electrolyte (see, {\it e.g.}, Ref.~\cite{Crocker96}), the electric force between colloids can be of the same order of magnitude as the external gravitational force, making this experimental setup feasible in relation to our theory.}

At this point, we would like to clarify some of the limitations of our framework. First, beyond the linear approximation, quadratic terms such as $c\nabla\psi$ appear in the Stokes and force balance equations, mixing these three decoupled equations.
Second, our framework was derived for a dilute electrolyte. Beyond the dilute limit, steric effects between the ions become important, and can be described by a lattice gas model. Including the corresponding pressure term in the formalism yields the modified Poisson-Nernst-Planck framework~\cite{Kilic07}.

In addition, ionic correlations also become important for high ionic concentrations. They affect not only the pressure, but  also the solvent viscosity~\cite{JD,Falkenhagen32,Onsager32} and the dielectric constant~\cite{Onsager,Kirkwood,Hasted48}.
However, we note that within our linear framework, such a dependence on the ionic concentration amounts to a mere shift in the values of the dielectric constant and viscosity, corresponding to the homogeneous value $n_0$. Consequently, all the results derived in this work remain unchanged.

As a possible future extension of our model, we recall that the independent response to the externally imposed chemical potential, $\mu$, and charge density, $q$, holds only for anions and cations with equal friction coefficients, as considered here. In principle, cations and anions can have different friction coefficients, $\zeta_+\ne\zeta_-$, due to their different sizes and chemical properties~\cite{Vanysek}.
Such a friction asymmetry couples between the number density and electric variables. For example, an external charge density, $q$, imposes oppositely directed forces on the cations and anions. Due to their different mobilities, a number density current is generated with its corresponding number density. The response functions for asymmetric friction coefficients are derived in  Appendix~\ref{appB}, and will be investigated further in a future work.

Another natural extension of our linear-response theory would be to include boundaries, such as flat surfaces. The generalization of the Oseen tensor for a flat surface with a no-slip boundary condition is known as the Blake tensor~\cite{Blake}. Our result for the electrostatic potential in Eq.~(\ref{eq27}) can be similarly modified for a flat boundary condition by the use of the ``image charge'' method. This scenario will be explored elsewhere.

%Finally, we would like to highlight possible applications of our findings.
As was mentioned before, the linear response to moving sources is relevant for several colloidal systems. Examples vary from sedimenting colloids, colloidal probe AFM microscopy or manipulation of optically trapped colloids in solution. In addition, this work can be of use in biological systems, where charged molecules interact in aqueous environments under flow.
For example, it was shown that charged biomolecules are only partially screened due to the presence of electro-diffusion current flow in aqueous pores~\cite{Liu08}.
We hope that our general framework can further applied in a wide range of physical and biological systems.

%%%%%%%%%%
\acknowledgments
%%%%%%%%%%

We thank H. Diamant, \raa{M. Urbakh, and M. Bazant} for fruitful discussions and helpful suggestions.
RA thanks the hospitality of Tokyo Metropolitan University, where part of this research was
conducted under the TAU-TMU co-tutorial program.
YU was supported by a Grant-in-Aid for the Japan Society for the Promotion of Science (JSPS)
Research Fellow No.\ 16J00042.
He also thanks the hospitality of Tel Aviv University, where this work has been completed.
SK acknowledges support by a Grant-in-Aid for Scientific Research (C) (Grant No.\ 18K03567) from
the JSPS.
DA thanks the ISF-NSFC joint research program under Grant No.\ 885/15 for partial support.

%%%%%%%%%%%%%%%%%%%%%%%%%%%%%%%%%%%%%%%%%%%%%%%%%
\appendix
\section{Linearization scheme}
\label{appA}

Consider a small perturbation to the homogenous solution due to three weak sources: an external force density, $\boldsymbol{F}$, a charge density, $q$, and an externally imposed chemical potential, $\mu$. The resulting fields, which have been rescaled in Eq.~(\ref{eq9}), can be expanded around the homogeneous solution as
%%%%%%%%%
\begin{align}
\label{eqa1}
&\bu \approx \buo+\bu_1,~~~~~P\approx 1+P_1 \, , \nonumber\\
&n\approx1+n_1,~~~~~\bjn\approx \bjn_1 \, ,\nonumber\\
&c\approx c_1,~~~~~\bjc\approx\bjc_1 \, ,~~~~~\psi\approx \psi_1 \, .
\end{align}
%%%%%%%%%%%%%%%%
The linear corrections, denoted by the subscript ``1'', are considered to be of a small magnitude, comparable to that of the sources.

Expanding Eq.~(\ref{eq10}) to first order leads to the following set of seven equations
corresponding to Eqs.~(\ref{eq11})--(\ref{eq13}):
%%%%%%%%%%%%%%
\begin{align}
\label{eqa2}
\eta\nabla^2\bu_1(\vecr)-\nabla P_1(\vecr)&=
-\boldsymbol{F}(\vecr) \, ,\nonumber\\
\nabla^2\psi_1(\vecr)+c_1(\vecr)&=
-q(\vecr) \, ,\nonumber\\
\bjn_1(\vecr)+\nabla n_1(\vecr)&=
-\nabla\mu(\vecr) \, ,\nonumber\\
\bjc_1(\vecr)+\nabla c_1(\vecr)
&=-\nabla \psi_1(\vecr) \, ,\nonumber\\
\nabla \cdot \bu_1(\vecr)&=0 \, ,\nonumber\\
\nabla\cdot\bjn_1(\vecr)
+\buo\cdot\nabla n_1(\vecr)&=0 \, ,\nonumber\\
\nabla\cdot\bjc_1(\vecr)
+\buo\cdot\nabla c_1(\vecr)&=0 \, .
\end{align}
%%%%%%%%%%%%%%
Because all the coupling terms are of second order in our correction fields, the above equations are decoupled. The fluid velocity $\bu_1$ and pressure $P_1$ depend solely on $\bF$. The number density $n_1$ and corresponding current $\bjn_1$ depend solely on $\mu$.
The charge density $c_1$ and electric current $\bjc_1$ depend solely on $q$. The potential $\psi$ is related to the charge density source $q$ via Poisson's equation.

 In  Fourier space, Eq.~(\ref{eqa2}) becomes
%%%%%%%%%%%%%%
\begin{align}
\label{eqa3}
-\eta k^2\bu_1(\veck)-i\veck P_1(\veck)&=
-\boldsymbol{F}(\veck) \, ,\nonumber\\
k^2\psi_1(\veck)-c_1(\veck)&=
-q(\veck)\, ,\nonumber\\
\bjn_1(\veck)+i\veck n_1(\veck)&=
-i\veck \mu(\veck) \, ,\nonumber\\
\bjc_1(\veck)+i\veck c_1(\veck)
&=-i\veck \psi_1(\veck) \, ,\nonumber\\
i\veck \cdot \bu_1(\veck)&= 0 \, ,\nonumber\\
 i\veck\cdot\bjn_1(\veck)
 +i\buo\cdot\veck n_1(\veck)&=0 \, ,\nonumber\\
i\veck\cdot\bjc_1(\veck)
+i\buo\cdot\veck c_1(\veck)&=0 \, .
\end{align}
%%%%%%%%%%%%%%
This is a set of linear equations to be solved in Fourier space as
%%%%%%%%%%%%%
\begin{align}
\label{eqa4}
P_1(\veck)&= -\frac{i}{k^2}\veck\cdot\bF(\veck) \, ,\nonumber\\
\bu_1(\veck) &=\frac{1}{\eta k^2}\left(\mathbf{I}
-\frac{{\veck\veck}}{k^2}\right)\cdot \bF(\veck) \, ,\nonumber\\
n_1(\veck)&=-\frac{k^2}{k^2+i\buo\cdot\veck}\,\mu(\veck) \, ,\nonumber\\
\bjn_1(\veck)&=\frac{(\buo\cdot\veck)\veck}{k^2
+i\buo\cdot\veck}\,\mu(\veck) \, ,\nonumber\\
\psi_1(\veck)&=\frac{k^2
+i\buo\cdot\veck}{k^2(1+i\buo\cdot\veck+k^2)}\, q(\veck) \, ,\nonumber\\
c_1(\veck)&=
-\frac{k^2}{k^2+i\buo\cdot\veck}\,\psi_1(\veck) \, ,\nonumber\\
\bjc_1(\veck)&=\frac{(\buo\cdot\veck)\veck}{k^2+i\buo\cdot\veck}\,\psi_1(\veck) \, .
\end{align}
%%%%%%%%%%%%%%
The above solution is written in Eqs.~(\ref{eq15}), (\ref{eq17}), (\ref{eq25}) and (\ref{eq26}), and the real-space solutions can be obtained by taking the appropriate convolutions.

%%%%%%%%%%%%%%%%%%%%%%
\section{Asymmetric friction coefficients}
\label{appB}

Generally, cations and anions may have different friction coefficients, $\zeta_{\pm}$.
Such a friction asymmetry results in a slightly modified set of equations. The force balance equations for the ions
in Eq.~(\ref{eq6}) is now replaced by
%%%%%%%%%%%%%%%%%%%%%%%%%%%%%%%%%%%%%%
\begin{align}
\label{eqb1}
- en_{+}\nabla \psi - \zeta_+ n_{+}\left(\boldsymbol{v}_{+}-\boldsymbol{u}\right)-\kbt\nabla n_{+} & =n_{+}\nabla \mu \, ,\nonumber\\
e n_{-}\nabla \psi -
 \zeta_- n_{-}\left(\boldsymbol{v}_{-}-\boldsymbol{u}\right)-\kbt\nabla n_{-} & =n_{-}\nabla \mu \, .
\end{align}
%%%%%%%%%%%%%%%%%%%%%%%%%%%%%%%%%%%%%%%%%%%
Note that the original physical quantities (and not dimensionless ones) are used in the above equation. Following the same scheme as in Sec.~\ref{ssec2a}, it is possible to define the dimensionless variables for the number density $n$ and charge density $c$. In the scaling of the velocities [see Eq.~(\ref{eq9})], we use the average friction coefficient, $\zeta=(\zeta_{+}+\zeta_{-})/2$. In addition, we define the dimensionless friction asymmetry parameter:
%%%%%%%%%
\begin{align}
\label{eqb2}
\xi=\frac{\zeta_{+}-\zeta_{-}}{\zeta_{+} +\zeta_{-}} \, .
\end{align}
%%%%%%%%

The above parameter  $\xi$ couples between the number density and charge density fields. Explicitly, Eq.~(\ref{eq10}) is replaced by the following set of seven equations:
%%%%%%%%%%%%
\begin{align}
\label{eqb3}
\eta\nabla^{2}\boldsymbol{u}-\nabla P
- c\nabla \psi&=-\boldsymbol{F} \, ,\nonumber \\
\nabla^{2}\psi+c&=-q \, ,\nonumber \\ \boldsymbol{J}+\xi\bjc+c\nabla \psi+\nabla n&=-n\nabla\mu \, ,\nonumber \\
\boldsymbol{j}+\xi\bjn+n\nabla \psi+\nabla c&=-c\nabla\mu \, ,\nonumber \\
\nabla \cdot\boldsymbol{u}&=0 \, ,\nonumber \\
\nabla \cdot\boldsymbol{J}+\bu\cdot\nabla n&=0 \, ,\nonumber \\
\nabla \cdot\boldsymbol{j}+\bu\cdot\nabla c&=0 \, .
\end{align}
%%%%%%%%%%%%%%
While most of the equations remain unchanged, the third and fourth equations contain a new term that
is linear in $\xi$. It is evident that Eq.~(\ref{eq10}) is restored for $\xi=0$.

Employing the same linearization scheme, the linear correction terms can be obtained in Fourier space. The hydrodynamic fields, $P_1$ and $\bu_1$, are given by the Oseen's result in Eq.~(\ref{eq15}). The number density and charge density fields are now given by
%%%%%%%%%%%%%%%%5
\begin{align}
\label{eqb4}
n_1(\veck)&=A(\veck)\,\mu(\veck)+\xi B(\veck)\,\psi_1(\veck)  \, ,\nonumber\\
\bjn_1(\veck)&=-(\buo\cdot\hat{\veck})\hat{\veck}\,n_1(\veck)\, ,\nonumber\\
c_1(\veck)&=\xi B(\veck)\,\mu(\veck)+A(\veck)\,\psi_1(\veck)\, ,\nonumber\\
\bjc_1(\veck)&=-(\buo\cdot\hat{\veck})\hat{\veck}\,c_1(\veck) \, ,
\end{align}
%%%%%%%%%%%%%%%%%%%%%%%%
where the functions $A(\veck)$ and $B(\veck)$ are defined as
%%%%%%%%%%%%%%%%
\begin{align}
\label{eqb5}
A(\veck)&=-\frac{k^2(k^2+i\buo\cdot\veck)}{k^2+(i\buo\cdot\veck)^2+(\xi\buo\cdot\veck)^2} \, ,\nonumber\\
B(\veck)&=\frac{k^2(i\buo\cdot\veck)}{(k^2+i\buo\cdot\veck)^2+(\xi\buo\cdot\veck)^2} \, .
\end{align}

%%%%%%%%%%%%%%%%%%%%%%%%
Finally, the electrostatic potential $\psi_1$ is related to $q$ and $\mu$ by the relation
%%%%%%%%%%%%%%%%
\begin{align}
\label{eqb6}
\psi_1(\veck)=\frac{q(\veck)+\xi B(\veck)\mu(\veck)}{k^2[1-A(\veck)]} \, .
\end{align}
%%%%%%%%%%%%%%

Equations~(\ref{eqb4}) and (\ref{eqb5}) are the generalizations of Eqs.~(\ref{eq17}) and (\ref{eq26}), respectively,  for asymmetric friction coefficients, $\zeta_+\ne\zeta_-$. In both cases, the currents $\bjn$ and $\bjc$ are related to $n$ and $c$, respectively, by a factor of $-(\buo\cdot\hat{\veck})\hat{\veck}$, as given by the corresponding continuity equations. The response of the $n$ and $\bjn$ fields to their direct source $\mu$ and indirect one $\psi_1$ is found to be the same as the response of $c$ and $\bjc$ to their direct source $\psi_1$ and indirect one $\mu$.

%\newpage
%%%%%%%%%%%%%%%

\end{document}